\documentclass[aps,prl,superscriptaddress]{revtex4-2}
\usepackage{amsmath,amsfonts,amssymb,verbatim,mathtools}
\usepackage{amsfonts}
\usepackage{graphicx}
\usepackage[normalem]{ulem}
\usepackage{color}

\newcommand{\strike}[1]{{\color{red}\sout{#1}}}
\renewcommand{\strike}[1]{}

\usepackage{dcolumn}
\usepackage{tabularx}
\usepackage{keyval}
\usepackage{multirow}

\newcommand{\BI}{(Pb,Bi)$_{2}$Sr$_{2-x}$La$_x$CuO$_{6+\delta}$ }

\newcommand{\WZI}{Van der Waals - Zeeman Institute, Institute of Physics, University of Amsterdam, Sciencepark 904, 1098 XH Amsterdam, The Netherlands}
\newcommand{\UU}{Institute for Theoretical Physics and Center for Extreme Matter and Emergent Phenomena, Utrecht University, The Netherlands}
\newcommand{\DIEP}{Dutch Institute for Emergent Phenomena (DIEP), Sciencepark 904, 1098 XH Amsterdam, The Netherlands}
\newcommand{\DLS}{Diamond Light Source, Harwell Campus, Didcot OX11 0DE, United Kingdom}
\newcommand{\Leiden}{Institute-Lorentz for Theoretical Physics, Leiden University, P.O. Box 9506, Leiden, Netherlands.}
\newcommand{\nordita}{NORDITA, KTH Royal Institute of Technology and Stockholm University, Hannes Alfv\'ens v\"ag 12, 106 91 Stockholm, Sweden}
\newcommand{\tokyo}{Institute for Solid State Physics, University of Tokyo, Kashiwa, Chiba 277-8581, Japan}
\newcommand{\nagoya}{Energy Materials Laboratory, Toyota Technological Institute 2-12-1 Hisakata Tempaku-ku, Nagoya 468-8511, Japan}
\newcommand{\radboud}{High Field Magnet Laboratory (HFML-EMFL) and Institute for Molecules and Materials, Radboud University, Toernooiveld 7, 6525 ED Nijmegen, The Netherlands}
\newcommand{\bristol}{H. H. Wills Physics Laboratory, University of Bristol, Tyndall Avenue, Bristol BS8 1TL, United Kingdom}

\begin{document}
\title{Momentum-dependent scaling exponents of nodal self-energies measured in strange metal cuprates and modelled using semi-holography}

\author{S. Smit}\email{s.smit@uva.nl}
\affiliation{\WZI}

\author{E. Mauri}
\affiliation{\UU}

\author{L. Bawden}
\affiliation{\WZI}

\author{F. Heringa}
\affiliation{\WZI}

\author{F. Gerritsen}
\affiliation{\WZI}

\author{E. van Heumen}
\affiliation{\WZI}

\author{Y.K. Huang}
\affiliation{\WZI}

\author{T. Kondo}
\affiliation{\tokyo}

\author{T. Takeuchi}
\affiliation{\nagoya}

\author{N. E. Hussey}
\affiliation{\radboud}
\affiliation{\bristol}

\author{T.K. Kim}
\affiliation{\DLS}

\author{C. Cacho}
\affiliation{\DLS}

\author{A. Krikun}
\affiliation{\nordita}
\author{K. Schalm}
\affiliation{\Leiden}

\author{H.T.C. Stoof}
\affiliation{\UU}

\author{M.S. Golden}\email{m.s.golden@uva.nl}
\affiliation{\WZI}
\affiliation{\DIEP}

\begin{abstract} 
{ The anomalous strange metal phase found in high-$T_c$ cuprates does not follow the conventional condensed-matter principles enshrined in the Fermi liquid and presents a great challenge for theory. Highly precise experimental determination of the electronic self-energy can provide a test bed for theoretical models of strange metals, and angle-resolved photoemission can provide this as a function of frequency, momentum, temperature and doping. Here we show that constant energy cuts through the nodal spectral function in \BI  have a non-Lorentzian lineshape, meaning the nodal self-energy is $k$ dependent. We show that the experimental data are captured remarkably well by a power law with a $k$-dependent scaling exponent smoothly evolving with doping, a description that emerges naturally from AdS/CFT-based semi-holography. This puts a spotlight on holographic methods for the quantitative modelling of strongly interacting quantum materials like the cuprate strange metals.}

\end{abstract}

\maketitle

The rich temperature versus doping diagram of the cuprate high-$T_c$ superconductors presents a cornucopia of non-conformity \cite{Keimer2015}.
In their normal state, these unusual metals have qualitatively different experimental behaviour than the Fermi liquid, with in particular the linear-in-$T$ resistivity and its dissonant twin, the quadratic-in-$T$ Hall angle, being among the most strange. 

One intruiging possible explanation of strange metal phenomenology is that it is controlled by a strongly interacting quantum critical {\em phase}.
If so, the immediate question follows whether their properties can be described by holographic emergence principles, as these approaches are uniquely suitable for describing such physics.
Evidence for a quantum critical phase comes from high magnetic-field electrical transport experiments on overdoped cuprates \cite{cooper2009, Ayres2021}, and from laser-based Angle-Resolved PhotoEmission Spectroscopy (ARPES) on Bi$_{2}$Sr$_{2}$CaCu$_{2}$O$_{8+\delta}$ (Bi-2212), in which the normal state self-energy of the electrons in the nodal direction not only showed power-law behaviour in both $\omega$ and $T$, but the power varied smoothly from less than unity for underdoping, to the marginal Fermi-liquid value of unity for optimal doping \cite{Varma1989} (a result first reported in 1999 \cite{Valla1999}), to $1.2$ for overdoping such that $T_c \simeq 0.75 T_c^{max}$ \cite{Reber2019}.

Here, we explore the normal state over a much larger frequency and temperature range than before by studying overdoped samples of the single-CuO$_2$-plane cuprate \BI (Pb,Bi)-2201, using high-resolution ARPES.
We uncover a {\em qualitatively} new facet of the self-energy of the nodal electrons, namely that it is not only a function of $\omega$ and $T$, but also dependent on the magnitude of the momentum away from the Fermi momentum $k-k_{F}$, yielding constant energy cuts through the spectral function (MDCs) with a non-Lorentzian lineshape.

After presenting the experimental data, we switch gears to test whether a quantitative theoretical description can be found using the tools of the holographic duality known as the Anti-de Sitter/Conformal Field Theory correspondence (AdS/CFT) from string theory \cite{hartnoll2016holographic, Zaanen2015}.
AdS/CFT approaches have been shown to successfully capture both Fermi-liquid-like \cite{Cubrovic2009} and non-Fermi-Liquid-like aspects of strange metallic behavior \cite{Faulkner2010}, including power-law self-energies with smoothly varying scaling exponents \cite{Liu2011, Leong2017}.

Adopting a semi-holographic theoretical treatment \cite{Faulkner:2010tq}, whose details are given in the Methods section, we can match its predicted $k$-dependent spectral function to the high-precision ARPES data across the doping range studied, for a wide energy range below the Fermi level $E_F$, and for momenta well away from $k_F$. 

The combination of our experimental and theoretical results show that the $k$ dependence of the exponents is not only experimentally detectable, but also quantifiable in sufficient detail to pose a real test for any theoretical explanation, one such theory being semi-holography.

What emerges is that, firstly, our ARPES data confirm that nodal electronic self-energies display a power law in frequency and temperature.
Secondly, the scaling exponents describing the self-energies are experimentally shown to be momentum dependent.
Thirdly, the observed exponents at $k_F$ grow smoothly from unity at optimal doping towards the Fermi-liquid value of two, though this is never reached, even for non-superconducting samples.
These results further strengthen the notion that the optimally and overdoped Bi-based strange metals do represent a novel quantum critical phase.

\subsection{Nodal ARPES data and power-law analysis with $k$-independent self-energy}

\begin{figure}[h!]
\includegraphics[width=17cm]{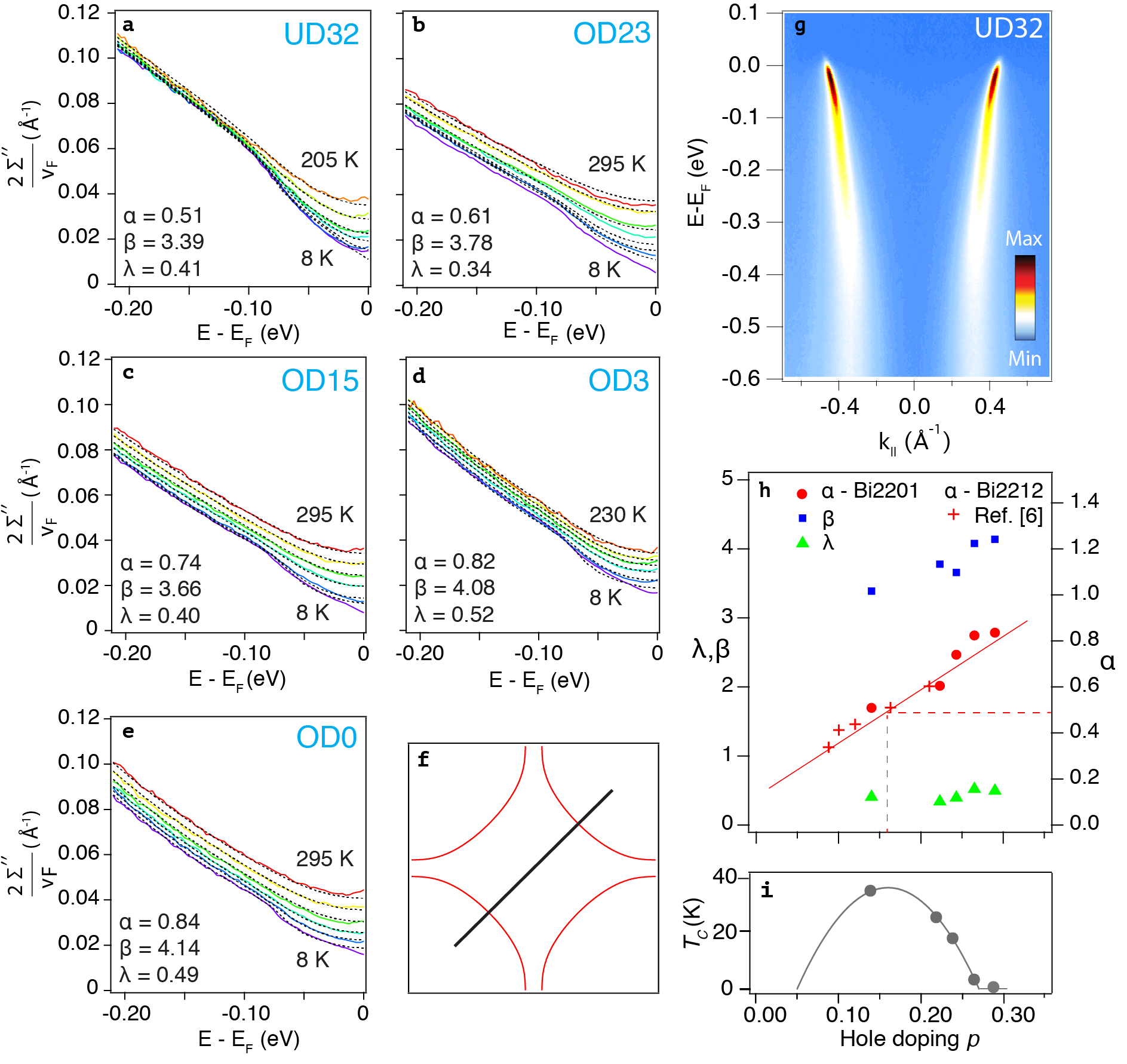}
\caption{ {\bf Nodal self-energy of single-CuO$_2$-layer (Pb,Bi)-2201.} {\bf a}-{\bf e}: Temperature-dependent self-energies from ARPES for five different doping levels, extracted using symmetric Voigt fits to MDCs, plotted using colour-coded solid lines for temperature. The dashed, black lines are results to two-dimensional (in $T$ and $\omega$) fits employing three parameters $(\alpha,\beta,\lambda)$, using the power-law-liquid formalism introduced in \cite{Reber2019} and given in Equation 1. Fitting parameters used are indicated in each data panel, and gathered together with the analogous parameters for Bi-2212 \cite{Reber2019} in panel {\bf h}. The red dashed lines indicate the marginal Fermi liquid ($\alpha$ = 0.5 at optimal doping), which is shown above the canonical Presland dome {\bf i} used to determine the doping level of the measured samples. Panel \textbf{g} shows a typical measured ARPES dataset, containing both the positive-$k$ and negative-$k$ nodal branches, along the $k$-space direction indicated in the schematic Fermi-surface in \textbf{f}.  Table S1 in the SI lists all temperatures measured for each doping level.}
\label{pll}
\end{figure}

In Fig. \ref{pll} we show the imaginary part of the nodal self-energy of \BI ~over a large range in frequency, doping and temperature. 
Assuming the bare dispersion to be linear, $\varepsilon({k}) = v_{F}({k}-k_{F})$, with $v_{F}$ being the Fermi velocity and $k_{F}$ the Fermi wave vector, the commonly made assumption of negligible $k$ dependence - $\Sigma({k},\omega) \simeq \Sigma(\omega)$ - reduces the single-particle spectral function to a Lorentzian lineshape as a function of momentum ${k}$ at each fixed frequency $\omega$, i.e., $L(k)= \frac{W}{\pi}\frac{\Gamma/2}{(k-k_{*})^{2}+(\Gamma/2)^{2}}$.
Here $W(\omega)$ is the intensity, $k_*(\omega)$ the peak position, and $\Gamma(\omega)$ its width.
For the results presented in the rest of this paper, $W(\omega)$ is not a key parameter and will not be discussed further.
In practice, the peaks are better fit using a Voigt lineshape, the Gaussian part of which accounts for experimental resolution. In this commonly adopted framework, the imaginary part of the self-energy is then proportional to the width of the Lorentzian component as $\Sigma^{''}(\omega) = v_{F}\Gamma(\omega)/2$ \cite{Valla1999}.
\\

Confirming and extending the results in \cite{Reber2019}, the ($\omega$,$T$)-dependent self-energy extracted in this way and shown in Fig.\ref{pll}
can be very well described by a remarkably simple phenomenological model dubbed the Power Law Liquid (PLL) \cite{Reber2019}, with three dimensionless parameters $\alpha$, $\beta$ and $\lambda$: 
\begin{equation}\label{Reber_pll}
\Gamma = \frac{2\Sigma^{''}(\omega, T)}{v_F} = G_{0}(\omega,T) + \lambda\frac{ [(\hbar \omega)^{2} + (\beta k_{B} T)^{2} ]^{\alpha}}{(\hbar \omega_{N})^{2 \alpha - 1}}~.
\end{equation}
Here $\lambda$ is a coupling constant describing the strength of the interaction, normalized to an energy scale $\hbar \omega_{N} = 0.5$eV for all dopings, and the parameter $\beta$ sets the balance between the relative influence of temperature and frequency.
$G_{0}(\omega,T)$ is an extra term, combining a self-energy contribution from impurity scattering, as well as electron-phonon coupling, described in the SI (S1). 
The electron-phonon contribution is most clearly seen around an energy of 70 meV \cite{Bogdanov2000,Cuk2005}, and is prominent because the normal state ARPES data presented here access much lower temperatures than those from Bi-2212 \cite{Reber2019}.   
Note that Equation~(\ref{Reber_pll}) describes the marginal Fermi-liquid \cite{Varma1989} with a power of unity ($\alpha = 1/2$) at optimal doping, and the quadratic temperature and frequency behaviour of a Fermi liquid emerges when $\alpha = 1$.

Figs.\ref{pll}b-f show the experimentally extracted self-energies (coloured lines), together with the result of a two-dimensional ($\omega$,$T$) fit to the data using Equation~(\ref{Reber_pll}) (dashed lines). 
An overview of the parameters extracted from the fits is given in Fig.~\ref{Reber_pll}g, together with the parameters for Bi-2212 \cite{Reber2019}. Compared to the latter, the ARPES data presented here cover a complementary doping range from near optimal doping ($p=0.14$) to such overdoping that superconductivity disappears.
Across this doping interval, the power increases smoothly from 1.04 ($\alpha=0.52$) to 1.68 ($\alpha= 0.84$), meaning that even on overdoping out of the superconducting dome, the quadratic power of the Fermi-liquid ($\alpha=1$) is not yet reached.

The continuity in $\alpha$ values evident in Fig.~\ref{Reber_pll}g emphasises the remarkably good agreement between the nodal self-energy behaviour in both single- and bi-layer cuprates, also illustrating that these self-energies near $E_F$ are well described by Equation ~\ref{Reber_pll} for a temperature range spanning 200K.

\subsection{Momentum-dependent power-law exponent and asymmetric ARPES MDCs}
\begin{figure}[h!]
\includegraphics[width=16cm]{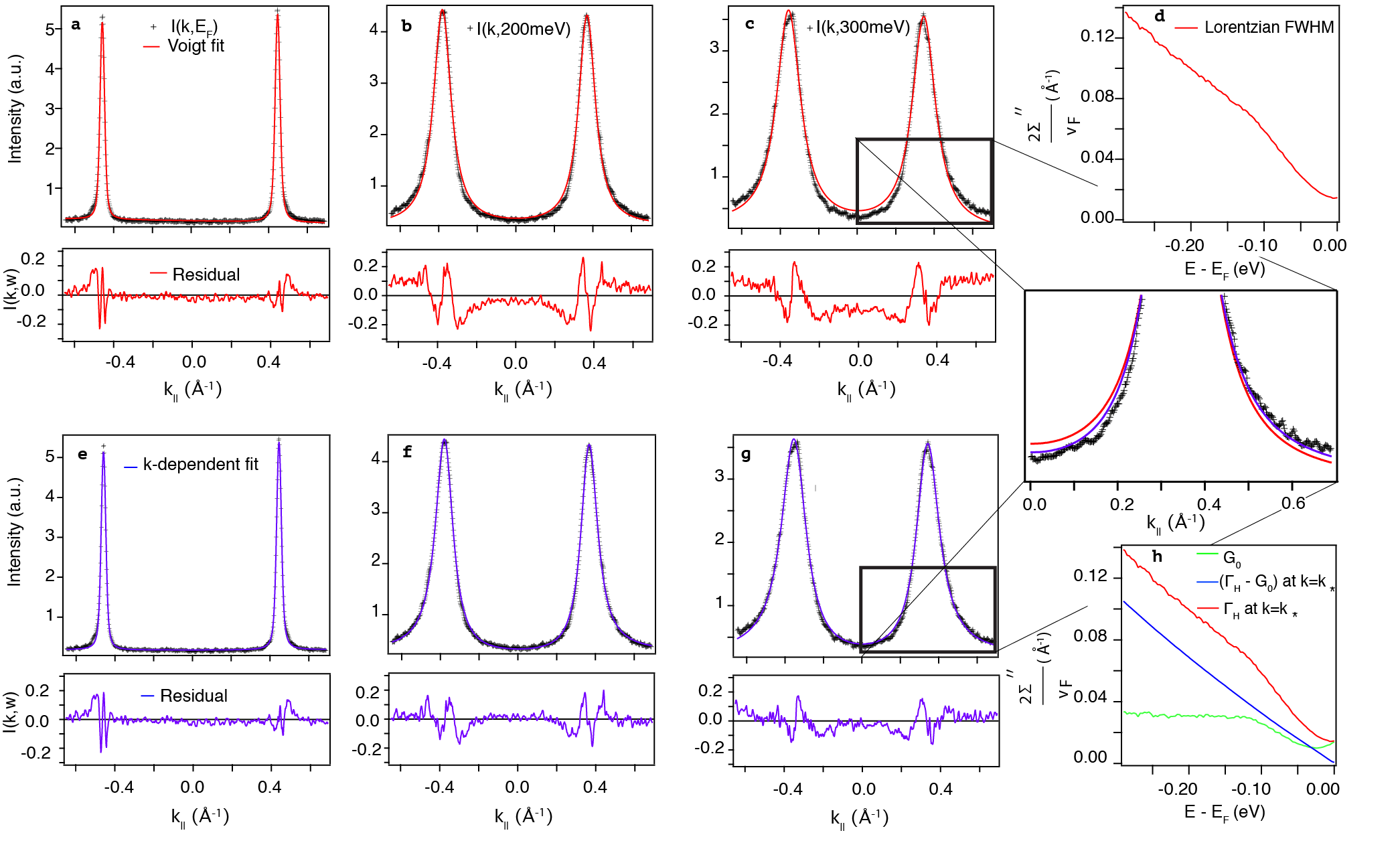}
\caption
{\textbf{Closer look at MDC lineshapes for UD32 nodal ARPES data.} \textbf{a}-\textbf{c}: a trio of MDCs at energies indicated, with symmetric peak fits in red. The residuals grow as the binding energy grows. Here \textbf{d} shows the resulting $2\Sigma''(\omega)/v_F = \Gamma(\omega)$ from the symmetric fits. \textbf{e}-\textbf{g}: The same three MDCs now fit in purple using our model given by Equation \ref{Mod_lor}, including the holographically-predicted $k$ dependence, with $V$ set to -1.
The residuals are clearly superior to the symmetric fit for energies further from $E_F$. \textbf{h:}  The imaginary part of the self-energy  $2\Sigma^{''}(k, \omega)/v_F = \Gamma_H(k, \omega)$ at $k_*(\omega)$ (red), which includes the $k$-dependent self-energy (blue), and the free fitting parameter $G_0$ (green).}
\label{MDC_fits}
\end{figure}

The above analysis does hinge on the aforementioned assumption of negligible $k$ dependence: such that the spectral function part of the MDC lineshape can be well fit using a symmetric Lorentzian for each peak \cite{Gaebel2004}.
In Fig. \ref{MDC_fits}a-d we show the Lorentzian fits in detail.
Panel a shows that a Lorentzian spectral function gives a very good fit \emph{close to $E_F$}, and yields a small residual. 
However, at higher binding energy, such as at 200 and 300meV, the foot of the peaks clearly show that the data (black crosses) are not captured completely by the two symmetric peaks in the fit (red lines).
The experimental bottom line is that the MDC peaks in Fig. \ref{MDC_fits}b \& \ref{MDC_fits}c are asymmetric: showing more spectral weight at large momenta $|{k}| > |k_{*}(\omega)|$, compared to $|{k}| < |k_{*}(\omega)|$ (leading to residual values of differing signs inside and outside the MDC peak pair).
This behaviour cannot be captured by a fit based on Lorentzians and the power law liquid.
The analysis and discussion covered by Figs. S2-S4 of the SI excludes that the MDC asymmetry comes from the non-zero curvature of an underlying `bare band', from inhomogeneous detector response, or from strongly structured background signals.
A simple yet profound possible explanation of the MDC asymmetry is then a $k$ dependency of the electronic self-energy itself.

Inclusion of a $k$-dependent self-energy, which is missing in the PLL ansatz of Equation \eqref{Reber_pll}, provides a significant experimental extension to the data-based fitness tests for aspiring theories for these materials.    One model-independent approach would be to linearly expand the measured self-energy around $k_{F}$, reported in the SI (Fig. S8), which significantly enlarges the set of fitting parameters, by means of an additional frequency and temperature dependent function.  
Instead, we can rely on the theoretical input provided by AdS/CFT, and adjust the fitting ansatz without {\em any} expansion in the number of parameters.

\subsection{Semi-holographic theoretical description}

Physically, a semi-holographic model describes an electron that interacts with a CFT (accounting for a quantum critical state deformed by non-zero $T$ and chemical potential) via linear coupling to a fermionic operator ${\cal O}$ that has a unique scaling dimension. As a result, the self-energy becomes proportional to the correlation function $\langle {\cal O O}^\dagger \rangle (k,\omega,T)$ in the CFT. The latter can be determined from the holographic dictionary \cite{Zaanen2015, hartnoll2016holographic}, and automatically inherits certain scaling properties from the `critical' CFT. To be applicable to the cuprate strange metals, the model must have a dynamical critical exponent $z \rightarrow \infty$ in order to recover the power law liquid when $k$ is close to $k_F$.
From this imposition of local quantum criticality, there follows a fundamental condition from the theory side that the scaling exponents \emph{have to be momentum dependent} \cite{Faulkner2011a}. 

Here, for the CFT we use an Einstein-Maxwell-Dilaton model of holography, specifically the Gubser-Rocha model \cite{Gubser2010, Gubser2012}, as it offers an analytical treatment of the gravitational spacetime.
In the long-wavelength limit, within the framework of an emergent particle-hole symmetry, our semi-holography model then gives at $T=0$:
\begin{subequations} \label{holo_spectralfunction}
\begin{eqnarray}
    \frac{2 \Sigma^{''}(k, \omega)}{v_F} &=& \lambda\frac{ [(\hbar \omega)^{2} ]^{\alpha(k)}}{(\hbar \omega_{N})^{2 \alpha(k) - 1}}~,\\
    \alpha(k) &=& \alpha \left( 1-\frac{{k}-k_{F}}{k_{F}} \right)~.
\end{eqnarray}
\end{subequations}

The semi-holographic self-energy can also be generalized to non-zero temperature \cite{Faulkner2011a, Gubser2012}, and under our conditions is well approximated by replacing $(\hbar \omega)^2$ with $(\hbar \omega)^2 + (\beta k_B T)^2$ in Equation (\ref{holo_spectralfunction}a).
This also eases comparison to the PLL in Equation (1), and highlights the key new insight that as frequency increases, and $k \simeq k_F - \omega/v_F$ departs from $k_F$, a $k$ dependence emerges in the \emph{exponent} describing the ($\omega,T$)-dependence of the self-energy. 
For low frequencies, where $k \simeq k_F$, one returns to Equation~\eqref{Reber_pll} of the PLL.
The behaviour predicted by holography should leave a clear experimental fingerprint, namely that the ARPES MDCs are asymmetric: the data shown in Fig. \ref{MDC_fits} show this is the case.

\begin{figure}[h!]
\includegraphics[width=9cm]{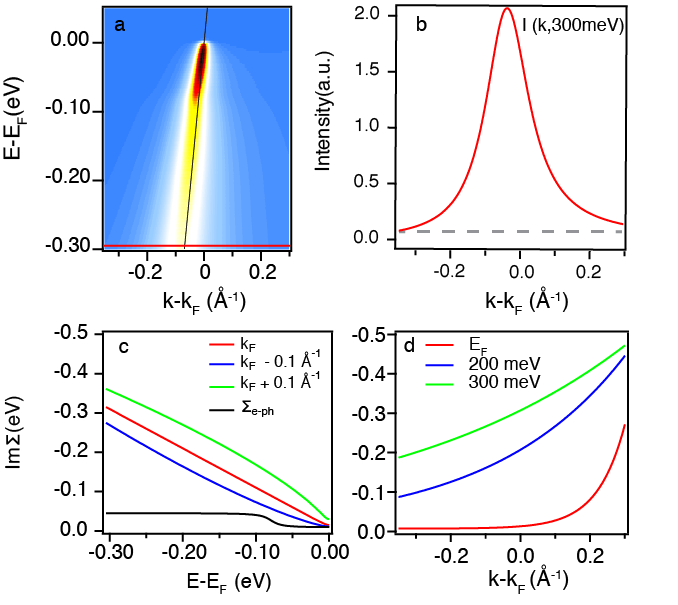}
\caption{{\bf Simulated spectral function at optimal doping, using the $k$-dependent self-energy from semi-holography.} \textbf{a}: Spectral function with a linear bare band (v$_F=$ 4 eV\AA, k$_{F}$=0.45\AA), convoluted with the experimental $k$ resolution and \textbf{b} MDC at non-zero binding energy generated using the full $\omega$ and $k$-dependent self-energy plotted in panels \textbf{c} and \textbf{d} from the semi-holographic model at $T$=20 K. Note that \textbf{c} includes the imaginary part of a $k$-independent phonon self-energy.}
\label{Holo_spectral}
\end{figure}

Fig.~\ref{Holo_spectral}a shows a simulated spectral function, generated using the semi-holographic self-energy shown in Figs.~\ref{Holo_spectral}c \& ~\ref{Holo_spectral}d.
The resulting MDC asymmetry at non-zero energy is illustrated in Fig.~\ref{Holo_spectral}b, visible as an increased intensity at the $|k|>|k_{*}|$ side of the peak maximum, just like in the experimental data of Fig. \ref{MDC_fits}c. 
The next step is to test whether this holographic approach can yield superior quantitative fits to the ARPES data.

\subsection{Fitting the semi-holographic theoretical model to the ARPES data}

As non-Lorentzian MDCs force a major departure from a cornerstone of ARPES data analysis methodology, we now describe how to deal with this in the analysis of real data from \BI.
The non-zero-$T$ version of the self-energy with the $k$-dependence in Equation (\ref{holo_spectralfunction}b), suggests a modified fit-function $L_{H}(k)$ at the fixed $\omega$ and $T$ relevant for each MDC:
\begin{subequations} \label{Mod_lor}
\begin{eqnarray} 
    L_{H}(k) &=& \frac{W}{\pi} \frac{\frac{\Gamma_H}{2}}{(k-k_{*})^2 + \left(\frac{\Gamma_H}{2}\right)^2}~,\\
    \Gamma_H(k) &=& G_{0} + \lambda \frac{[(\hbar\omega)^2+(\beta k_{B}T)^2]^{{\alpha({k})}}}{\hbar\omega_N^{2{\alpha({k})}-1}}~,\\
    \alpha({k})&=& \alpha \left(1 +V \left[\frac{{k}-k_{F}}{k_{F}}\right] \right)~.
\end{eqnarray}
\end{subequations}

$\Gamma_{H}(k)$ in Equation (\ref{Mod_lor}b) captures both the peak width and its asymmetry via momentum dependence built into the exponent $\alpha({k})$ in Equation (\ref{Mod_lor}c).
By fixing $\alpha$, $\beta$, $\lambda$ and $\omega_N$ to the PLL values at low energies, only $G_{0}$ and the asymmetry parameter $V$ remain free to vary in the fitting process for each MDC.

The Gubser-Rocha holographic model used here, however, actually $\emph{requires}$ that $V=-1$ for all frequencies. 
Thus, in this case the resulting fit function has exactly the same number of free parameters as the PLL, and is used to fit the experimental MDCs as shown using the purple lines in Fig. \ref{MDC_fits}e-g.

Comparing Figs. \ref{MDC_fits}b and \ref{MDC_fits}c (PLL, $k$-independent) with Figs. \ref{MDC_fits}f and \ref{MDC_fits}g ($k$-dependent scaling exponents), it is clear that on adopting Equation \eqref{Mod_lor}, the residuals in the panels f and g for energies well below $E_F$ drop, becoming as low as they were for the PLL at $E_F$.
In particular, the zoom to Figs. \ref{MDC_fits}c and \ref{MDC_fits}g illustrates clearly that the asymmetry of the MDC peaks is now captured almost perfectly. 
Fig. \ref{MDC_fits}h shows the total self-energy along the loci of the MDC peak-maxima in red, with the holographic, k-dependent part in blue, and the free fitting parameter G$_0 (\omega,T)$ in green. The latter can be seen to automatically take on the combined form of an offset (impurity scattering) plus a step function centered at the phonon energy of 70 meV. 
\vspace{5mm}

\subsection{Testing the semi-holographic prediction}

\begin{figure}[h!]
\includegraphics[width=14cm]{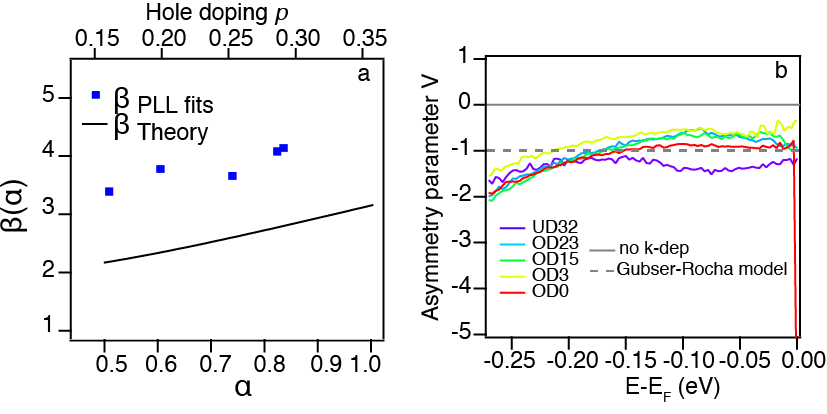}
\caption{{\bf Testing the predictions from semi-holography}
\textbf{a}: Comparison between the coefficient $\beta$, determining the amount of temperature dependence in the self-energy, obtained from the PLL fit to the ARPES data (blue squares) and predicted by the holographic model (black line), as a function of the doping-dependent power $\alpha$, which is translated into doping on the upper x-axis. {\bf b:} Asymmetry parameter $V$ of the momentum-dependent scaling exponent extracted from the experimental ARPES data over the full doping range measured at 8K, obtained by performing MDC fits to the data using Equation \eqref{Mod_lor}, where the leading scaling exponent $\alpha$, the coupling constant $\lambda$ and the quantity $(\hbar\omega)^{2} + (\beta k_{B} T)^{2}$ are fixed. A value $V = 0$ means there is no asymmetry, and $V = -1$ is the prediction from the semi-holographic Gubser-Rocha model. The high temperature fits are shown in the SI section S6.}
\label{doping_V}
\end{figure}

Upon determining all the free parameters in the theory (see Methods), two experimental tests of the semi-holographic results can then be performed. The first is a mild one, unconnected to the observed MDC asymmetry, and concerns the value of $\beta$, shown in Fig. \ref{doping_V}a. 
The experimental values (blue squares) all reside in an interval between $\beta =3-4$ across the doping levels studied.
Those coming from our model lie between $\beta =2-3$, and show a doping-dependent (generally upward) trend very similar to that of the experimental values. 
A detailed derivation of parameters such as $\beta$ falls outside the experimental focus of this paper, and will be presented in a separate publication. 

The second, more stringent test is given by the experimentally observed MDC asymmetry.
The non-zero value of $V$ is a non-trivial result, which is built into holography, but must be accounted for by other theoretical approaches aiming to describe the spectral function of the nodal charge carriers in cuprates. For the Gubser-Rocha model adopted here, the asymmetry parameter should have a frequency-independent value of $V=-1$.
We test this by performing a second fit of the ARPES data, now leaving $V$ as a free parameter for each $\omega$.
Fig. \ref{doping_V}b shows that - with no guiding restraint at all applied to $V(\omega)$ - the experimental data yield for all doping levels, temperatures (see SI) and frequencies a value of $V \approx -1$, close to the semi-holographic prediction.

Fig. \ref{doping_V} shows that there is still some room for improvement as regards to both $\beta$ and $V$ in the current approach.
Thus, these ARPES data and their parametrization presented here are an invitation to the many existing $z=\infty$ holographic models besides the analytical Gubser-Rocha model - or to completely different theoretical methods - to take the next step in capturing the observed $\beta$ values and modest frequency dependence of $V$ seen in experiment.

Before providing a discussion of other contexts in which k-dependent self energies have been discussed, we re-iterate the main results of this paper.
The $\omega$ and $T$ dependence of the electronic self-energy of nodal carriers in the normal state of the single-layer cuprate (Pb,Bi)-2201 is shown in ARPES experiments to follow a single power law with a $k$-\textit{dependent scaling exponent}, yielding asymmetric, non-Lorentzian MDCs for energies away from $E_F$. 
From fits to the ARPES data from optimal doping across the whole overdoped portion of the phase diagram that work well over wide ranges in energy and temperature of more then 200 meV and 250 K, respectively, at $k=k_F$ the $k$-dependent power-law exponent is found to take a nearly marginal Fermi-liquid value of unity ($\alpha = 0.51$) at optimal doping, growing to 1.68 ($\alpha = 0.84$) for the most overdoped, non-superconducting system studied. For $k=k_F$, these results connect smoothly to those of \cite{Reber2019}.
However, for values $k\neq k_F$ these powers change.
Such a $k$ dependence is a key prediction from (semi)holographic models of strange metals as a quantum critical phase.
The form of the $k$ dependence and the asymmetry parameter describing its magnitude are all successfully fitted using semi-holographic single-fermion spectral functions.
Importantly, any competing theoretical model for strange metals must be able to account for the observed ARPES MDC peak-asymmetry that is well described here by a momentum-dependent power-law self-energy.

There are few studies of the $k$ dependence of the self-energy in solids.
In the cuprate context, \cite{Randeria2004} and \cite{Veenstra2011} present theory/modelling of possible $k$ dependence in the self-energy resulting in a change in the dispersion velocity or arising from strong electron-phonon coupling, respectively. 
In a different condensed-matter realisation, $k$-dependent power laws have recently been observed in the zero-bias conduction anomaly in transport spectroscopy of nanowires \cite{Jin2019}, and linked to non-linear Luttinger liquid theory \cite{Imambekov2009}. 
One might therefore wonder if there is a link between the 1D physics in the nanowire case, and the carriers in the nodal $k$-space direction of the quasi-2D (Pb,Bi)-2201 system that seem well described by the 1D CFT encoded semi-holographically in the AdS$_2$ infrared geometry of the $z=\infty$ models.

The results presented here herald future comparisons of the experimentally-tested, theoretical spectral function to other experimental probes such as optical and DC conductivities \cite{Jacobs2015}. Alternatively, many-body condensed-matter theories can be guided by both the experimental self-energies presented here and the results from the AdS/CFT analogue, opening up possible pathways out of phenomenology and into a microscopic understanding of strange metals.

\section{Methods}

\subsection{ARPES measurements}

All ARPES data presented here were recorded at beamline I05 of Diamond Light Source, using (horizontally) linearly polarized light at a photon energy of 28 eV. 
Due to this chosen photon energy we can combine excellent energy resolution (12meV) with enough range in $k$ space to measure both the negative-$k$ and positive-${k}$ nodal branches in a single $I({k},\omega)$ image (see Fig. \ref{pll}A), which is very helpful in distinguishing the subtle effects at play from possible complicating factors such as signal background.
The energy resolution, together with the Fermi energy position - was confirmed by means of reference data from an amorphous Au film held in electrical contact with the sample.
To account for resolution in the angular direction, both the Lorentzian MDC fit function used for the PLL analysis, and the  L$_H$ MDC fit function from Equation\eqref{Mod_lor} are convolved with a Gaussian of width 0.01 \AA$^{-1}$. In the symmetric PLL case this yields a Voigt lineshape as indicated in the main text and in Fig. \ref{MDC_fits}.

All data were recorded in swept mode, to ensure any detector-response inhomogeneity is averaged out in the energy direction, and were also confirmed to be free of detector non-linearity effects \cite{Reber2014}. All data were recorded in the X$\Gamma$X direction in the Brillouin zone.

High quality single crystals of diffraction replica-free \BI, or (Pb,Bi)-2201), were grown using floating-zone techniques.
Individual crystals were annealed in varying atmospheres for varying lengths of time, so as to change the oxygen content and with it the hole doping, controlling the carrier concentration and $T_c$.
The critical temperatures were determined either via resistivity or AC-susceptibility measurements and the doping level was read off using $T_{c}$ and the Presland formula \cite{Presland1991, ando2000}. For the sample OD0K, $T_c\lesssim2K$ $\lesssim0.05 T_c^{max}$. 

\vspace{5mm}
\subsection{Holography and semi-holography}
The Anti-de Sitter/Conformal Field Theory correspondence (AdS/CFT) links a classical gravitational theory with one additional spatial dimension to a strongly interacting quantum theory on the boundary of the extended space. In practice, it provides a systematic way to compute the response functions for a quantum theory with strong interactions, even at non-zero temperature and in the presence of a chemical potential, by means of solving a (relatively speaking) simpler problem, namely solving classical gravitational equations.
For our purposes, therefore, the curved space is three dimensional and the quantum theory should describe the correlated physics of the maximally entangled fermions in the quasi-2D high-$T_c$ cuprates \cite{Zaanen2015, hartnoll2016holographic}. To be precise, holography describes the physics of a composite fermion $\cal O$ with a large number of degrees of freedom, i.e., in the so-called large-$N$ limit. However, as we here are interested in testing theoretical predictions against spectral functions obtained through ARPES data, we ultimately want to compute the response function for an elementary fermion $\psi$. The solution is found within the semi-holographic framework \cite{Faulkner:2010tq}, where the fermion $\psi$ is linearly coupled to the composite fermion $\cal O$ and the Green's function describing the dynamics of $\psi$ then acquires the form
\begin{equation}\label{green_semi}
  G_{\psi\psi}(\omega, k) = \frac{1}{\omega - v_F (k-k_F) - g_s G^{-1}(\omega, k)} \text{ ,}
\end{equation}
where $G^{-1}(\omega, k)$, assuming the role of the self-energy, is the two-point function of the composite fermionic operator $\cal O$, and $G_{\psi\psi}$ can be shown to be properly normalized as the Green's function of an elementary fermion \cite{Gursoy2012}. 
The most important piece of the puzzle, the qualitative form of the complex self-energy at energies near the Fermi level, thus comes from  holography. It is in this context that we compute the inverse Green's function $G^{-1}(\omega, k)$ appearing in Eq. \eqref{green_semi}. This is done by solving the Dirac equation on the curved gravitational background spacetime \cite{Iqbal2011}. 

In general, as it has been done for the model in this paper, the gravitational equations needed to compute the holographic Green's function can only be solved numerically. Nonetheless, the qualitative behavior of the low-energy Green's function of such a fermionic operator can be obtained analytically and it is completely determined by the infrared properties of the theory, captured in the duality by the inner geometry of the gravitational spacetime.
The electron self-energy obtained in this a way is the one we use for the nodal holes near the Fermi surface within the framework of an emergent particle-hole symmetry, and it possesses the desired ($\omega,T$)-scaling properties, i.e., power-law exponents matching the experimental ARPES data.
In particular, this implies that in the class of Einstein-Maxwell-Dilaton geometries characterized by the dynamical critical exponent $z = \infty$, and in the presence of a Fermi surface, the inverse Green's function at $T = 0$ (the low-energy behavior can be easily generalized to small non-zero temperature \cite{Gubser2012, Faulkner2011}) has to assume a low-energy form near $k_F$ \cite{hartnoll2016holographic} given by
\begin{equation}\label{green_composite}
  G^{-1}(\omega, k) = g \omega (-\omega^2)^{\nu_k-1/2} + \dots \text{ ,}
\end{equation}
where $g$ is constant in the limit $\omega, k-k_F \rightarrow 0$.
It is important to note that the $k$-dependent power $\nu_k$ is solely dictated by the IR of the theory and thus is independent of any of the UV details. The lifetime of the fermionic excitations near the Fermi surface are then described by
\begin{equation}
  \Sigma^{''}(\omega, k) = -\frac{\lambda v_F}{2} (\omega^2)^{\nu_k} + \dots~.
\end{equation}
Theoretically, the constant $\alpha \equiv \nu_{k_F}$ in the exponent in Eq. 2b, depends on the two (dimensionless) parameters in the Dirac equation, the mass $m$ and the charge $q$, that encode certain defining properties of the conformal field theory \cite{hartnoll2016holographic}. We keep the mass fixed close to the limit of $m = -1/2$ of the range allowed for by semi-holography ($-1/2 < m < 1/2$) \cite{Gursoy2012}. Then, varying only $q$ as a function of the hole doping $p$, captures the doping dependence of the exponent $\alpha(p)$ seen in the experimental data, using the numerically obtained linear relationship $\alpha \approx 1.9 q$. 
Finally, the value of $\lambda$ (see Eq. 2a) leading to a match with experiment at each doping level can be obtained by adjusting the strength of the coupling $g_s$ between the fermion $\psi$ and the conformal field theory.

We can see that Eq. \eqref{green_composite} has the same qualitative form as the semi-holographic response \eqref{green_semi} for $\omega, k-k_F \rightarrow 0$.
Semi-holography thus only changes the UV of the theory, i.e. the constant $g$, but it does not change the emergent behavior $\Sigma'' \propto (\omega^2)^{\nu_k}$. In this sense, while we showed that in our `bottom-up' approach we possess enough tuning knobs so as to get a semi-holographic spectral function that closely resembles the one observed in the ARPES experiments, we do not know whether the values used do indeed correspond to a consistent theory of gravity.
Nonetheless, our main point that the self-energy contains a momentum-dependent power is a universal feature of $z = \infty$ holographic calculations, independent of the particular UV completion applied.
In fact, while the theoretical model proposed here is rooted in holography, any quantum critical theory with infinite dynamical exponent and a large-$N$ suppression of higher-order correlation functions will be able to give rise to an effective correlation function of the same form as in Eq. \eqref{green_semi} \cite{hartnoll2016holographic}.

\newpage

\bibliography{ms}

%apsrev4-2.bst 2019-01-14 (MD) hand-edited version of apsrev4-1.bst
%Control: key (0)
%Control: author (8) initials jnrlst
%Control: editor formatted (1) identically to author
%Control: production of article title (0) allowed
%Control: page (0) single
%Control: year (1) truncated
%Control: production of eprint (0) enabled
\begin{thebibliography}{30}%
\makeatletter
\providecommand \@ifxundefined [1]{%
 \@ifx{#1\undefined}
}%
\providecommand \@ifnum [1]{%
 \ifnum #1\expandafter \@firstoftwo
 \else \expandafter \@secondoftwo
 \fi
}%
\providecommand \@ifx [1]{%
 \ifx #1\expandafter \@firstoftwo
 \else \expandafter \@secondoftwo
 \fi
}%
\providecommand \natexlab [1]{#1}%
\providecommand \enquote  [1]{``#1''}%
\providecommand \bibnamefont  [1]{#1}%
\providecommand \bibfnamefont [1]{#1}%
\providecommand \citenamefont [1]{#1}%
\providecommand \href@noop [0]{\@secondoftwo}%
\providecommand \href [0]{\begingroup \@sanitize@url \@href}%
\providecommand \@href[1]{\@@startlink{#1}\@@href}%
\providecommand \@@href[1]{\endgroup#1\@@endlink}%
\providecommand \@sanitize@url [0]{\catcode `\\12\catcode `\$12\catcode
  `\&12\catcode `\#12\catcode `\^12\catcode `\_12\catcode `\%12\relax}%
\providecommand \@@startlink[1]{}%
\providecommand \@@endlink[0]{}%
\providecommand \url  [0]{\begingroup\@sanitize@url \@url }%
\providecommand \@url [1]{\endgroup\@href {#1}{\urlprefix }}%
\providecommand \urlprefix  [0]{URL }%
\providecommand \Eprint [0]{\href }%
\providecommand \doibase [0]{https://doi.org/}%
\providecommand \selectlanguage [0]{\@gobble}%
\providecommand \bibinfo  [0]{\@secondoftwo}%
\providecommand \bibfield  [0]{\@secondoftwo}%
\providecommand \translation [1]{[#1]}%
\providecommand \BibitemOpen [0]{}%
\providecommand \bibitemStop [0]{}%
\providecommand \bibitemNoStop [0]{.\EOS\space}%
\providecommand \EOS [0]{\spacefactor3000\relax}%
\providecommand \BibitemShut  [1]{\csname bibitem#1\endcsname}%
\let\auto@bib@innerbib\@empty
%</preamble>
\bibitem [{\citenamefont {Keimer}\ \emph {et~al.}(2015)\citenamefont {Keimer},
  \citenamefont {Kivelson}, \citenamefont {Norman}, \citenamefont {Uchida},\
  and\ \citenamefont {Zaanen}}]{Keimer2015}%
  \BibitemOpen
  \bibfield  {author} {\bibinfo {author} {\bibfnamefont {B.}~\bibnamefont
  {Keimer}}, \bibinfo {author} {\bibfnamefont {S.~A.}\ \bibnamefont
  {Kivelson}}, \bibinfo {author} {\bibfnamefont {M.~R.}\ \bibnamefont
  {Norman}}, \bibinfo {author} {\bibfnamefont {S.}~\bibnamefont {Uchida}},\
  and\ \bibinfo {author} {\bibfnamefont {J.}~\bibnamefont {Zaanen}},\
  }\bibfield  {title} {\bibinfo {title} {{From quantum matter to
  high-temperature superconductivity in copper oxides}},\ }\href
  {https://doi.org/10.1038/nature14165} {\bibfield  {journal} {\bibinfo
  {journal} {Nature}\ }\textbf {\bibinfo {volume} {518}},\ \bibinfo {pages}
  {179} (\bibinfo {year} {2015})}\BibitemShut {NoStop}%
\bibitem [{\citenamefont {Cooper}\ \emph {et~al.}(2009)\citenamefont {Cooper},
  \citenamefont {Wang}, \citenamefont {Vignolle}, \citenamefont {Lipscombe},
  \citenamefont {Hayden}, \citenamefont {Tanabe}, \citenamefont {Adachi},
  \citenamefont {Koike}, \citenamefont {Nohara}, \citenamefont {Takagi},
  \citenamefont {Proust},\ and\ \citenamefont {Hussey}}]{cooper2009}%
  \BibitemOpen
  \bibfield  {author} {\bibinfo {author} {\bibfnamefont {R.~A.}\ \bibnamefont
  {Cooper}}, \bibinfo {author} {\bibfnamefont {Y.}~\bibnamefont {Wang}},
  \bibinfo {author} {\bibfnamefont {B.}~\bibnamefont {Vignolle}}, \bibinfo
  {author} {\bibfnamefont {O.~J.}\ \bibnamefont {Lipscombe}}, \bibinfo {author}
  {\bibfnamefont {S.~M.}\ \bibnamefont {Hayden}}, \bibinfo {author}
  {\bibfnamefont {Y.}~\bibnamefont {Tanabe}}, \bibinfo {author} {\bibfnamefont
  {T.}~\bibnamefont {Adachi}}, \bibinfo {author} {\bibfnamefont
  {Y.}~\bibnamefont {Koike}}, \bibinfo {author} {\bibfnamefont
  {M.}~\bibnamefont {Nohara}}, \bibinfo {author} {\bibfnamefont
  {H.}~\bibnamefont {Takagi}}, \bibinfo {author} {\bibfnamefont
  {C.}~\bibnamefont {Proust}},\ and\ \bibinfo {author} {\bibfnamefont {N.~E.}\
  \bibnamefont {Hussey}},\ }\bibfield  {title} {\bibinfo {title} {{Anomalous
  criticality in the electrical resistivity of La$_{2-x}$Sr$_x$CuO$_4$}},\
  }\href {https://doi.org/10.1126/science.1165015} {\bibfield  {journal}
  {\bibinfo  {journal} {Science}\ }\textbf {\bibinfo {volume} {323}},\ \bibinfo
  {pages} {603} (\bibinfo {year} {2009})}\BibitemShut {NoStop}%
\bibitem [{\citenamefont {Ayres}\ \emph {et~al.}(2021)\citenamefont {Ayres},
  \citenamefont {Berben}, \citenamefont {{\v{C}}ulo}, \citenamefont {Hsu},
  \citenamefont {van Heumen}, \citenamefont {Huang}, \citenamefont {Zaanen},
  \citenamefont {Kondo}, \citenamefont {Takeuchi}, \citenamefont {Cooper},
  \citenamefont {Putzke}, \citenamefont {Friedemann}, \citenamefont
  {Carrington},\ and\ \citenamefont {Hussey}}]{Ayres2021}%
  \BibitemOpen
  \bibfield  {author} {\bibinfo {author} {\bibfnamefont {J.}~\bibnamefont
  {Ayres}}, \bibinfo {author} {\bibfnamefont {M.}~\bibnamefont {Berben}},
  \bibinfo {author} {\bibfnamefont {M.}~\bibnamefont {{\v{C}}ulo}}, \bibinfo
  {author} {\bibfnamefont {Y.~T.}\ \bibnamefont {Hsu}}, \bibinfo {author}
  {\bibfnamefont {E.}~\bibnamefont {van Heumen}}, \bibinfo {author}
  {\bibfnamefont {Y.}~\bibnamefont {Huang}}, \bibinfo {author} {\bibfnamefont
  {J.}~\bibnamefont {Zaanen}}, \bibinfo {author} {\bibfnamefont
  {T.}~\bibnamefont {Kondo}}, \bibinfo {author} {\bibfnamefont
  {T.}~\bibnamefont {Takeuchi}}, \bibinfo {author} {\bibfnamefont {J.~R.}\
  \bibnamefont {Cooper}}, \bibinfo {author} {\bibfnamefont {C.}~\bibnamefont
  {Putzke}}, \bibinfo {author} {\bibfnamefont {S.}~\bibnamefont {Friedemann}},
  \bibinfo {author} {\bibfnamefont {A.}~\bibnamefont {Carrington}},\ and\
  \bibinfo {author} {\bibfnamefont {N.~E.}\ \bibnamefont {Hussey}},\ }\bibfield
   {title} {\bibinfo {title} {{Incoherent transport across the strange-metal
  regime of overdoped cuprates}},\ }\href@noop {} {\bibfield  {journal}
  {\bibinfo  {journal} {Nature}\ }\textbf {\bibinfo {volume} {595}},\ \bibinfo
  {pages} {661} (\bibinfo {year} {2021})}\BibitemShut {NoStop}%
\bibitem [{\citenamefont {Varma}\ \emph {et~al.}(1989)\citenamefont {Varma},
  \citenamefont {Littlewood}, \citenamefont {Schmitt-Rink}, \citenamefont
  {Abrahams},\ and\ \citenamefont {Ruckenstein}}]{Varma1989}%
  \BibitemOpen
  \bibfield  {author} {\bibinfo {author} {\bibfnamefont {C.~M.}\ \bibnamefont
  {Varma}}, \bibinfo {author} {\bibfnamefont {P.~B.}\ \bibnamefont
  {Littlewood}}, \bibinfo {author} {\bibfnamefont {S.}~\bibnamefont
  {Schmitt-Rink}}, \bibinfo {author} {\bibfnamefont {E.}~\bibnamefont
  {Abrahams}},\ and\ \bibinfo {author} {\bibfnamefont {A.~E.}\ \bibnamefont
  {Ruckenstein}},\ }\bibfield  {title} {\bibinfo {title} {{Phenomenology of the
  Normal State of Cu-O High-Temperature Superconductors}},\ }\href@noop {}
  {\bibfield  {journal} {\bibinfo  {journal} {Physical Review Letters}\
  }\textbf {\bibinfo {volume} {63}},\ \bibinfo {pages} {1996} (\bibinfo {year}
  {1989})}\BibitemShut {NoStop}%
\bibitem [{\citenamefont {Valla}\ \emph {et~al.}(1999)\citenamefont {Valla},
  \citenamefont {Fedorov}, \citenamefont {Johnson}, \citenamefont {Wells},
  \citenamefont {Hulbert}, \citenamefont {Li}, \citenamefont {Gu},\ and\
  \citenamefont {Koshizuka}}]{Valla1999}%
  \BibitemOpen
  \bibfield  {author} {\bibinfo {author} {\bibfnamefont {T.}~\bibnamefont
  {Valla}}, \bibinfo {author} {\bibfnamefont {A.~V.}\ \bibnamefont {Fedorov}},
  \bibinfo {author} {\bibfnamefont {P.~D.}\ \bibnamefont {Johnson}}, \bibinfo
  {author} {\bibfnamefont {B.~O.}\ \bibnamefont {Wells}}, \bibinfo {author}
  {\bibfnamefont {S.~L.}\ \bibnamefont {Hulbert}}, \bibinfo {author}
  {\bibfnamefont {Q.}~\bibnamefont {Li}}, \bibinfo {author} {\bibfnamefont
  {G.~D.}\ \bibnamefont {Gu}},\ and\ \bibinfo {author} {\bibfnamefont
  {N.}~\bibnamefont {Koshizuka}},\ }\bibfield  {title} {\bibinfo {title}
  {{Evidence for quantum critical behaviour in the optimally doped cuprate
  Bi$_{2}$Sr$_{2}$CaCu$_{2}$O$_{8+ \delta}$}},\ }\href@noop {} {\bibfield
  {journal} {\bibinfo  {journal} {Science}\ }\textbf {\bibinfo {volume}
  {285}},\ \bibinfo {pages} {2110} (\bibinfo {year} {1999})}\BibitemShut
  {NoStop}%
\bibitem [{\citenamefont {Reber}\ \emph {et~al.}(2019)\citenamefont {Reber},
  \citenamefont {Zhou}, \citenamefont {Plumb}, \citenamefont {Parham},
  \citenamefont {Waugh}, \citenamefont {Cao}, \citenamefont {Sun},
  \citenamefont {Li}, \citenamefont {Wang}, \citenamefont {Wen}, \citenamefont
  {Xu}, \citenamefont {Gu}, \citenamefont {Yoshida}, \citenamefont {Eisaki},
  \citenamefont {Arnold},\ and\ \citenamefont {Dessau}}]{Reber2019}%
  \BibitemOpen
  \bibfield  {author} {\bibinfo {author} {\bibfnamefont {T.~J.}\ \bibnamefont
  {Reber}}, \bibinfo {author} {\bibfnamefont {X.}~\bibnamefont {Zhou}},
  \bibinfo {author} {\bibfnamefont {N.~C.}\ \bibnamefont {Plumb}}, \bibinfo
  {author} {\bibfnamefont {S.}~\bibnamefont {Parham}}, \bibinfo {author}
  {\bibfnamefont {J.~A.}\ \bibnamefont {Waugh}}, \bibinfo {author}
  {\bibfnamefont {Y.}~\bibnamefont {Cao}}, \bibinfo {author} {\bibfnamefont
  {Z.}~\bibnamefont {Sun}}, \bibinfo {author} {\bibfnamefont {H.}~\bibnamefont
  {Li}}, \bibinfo {author} {\bibfnamefont {Q.}~\bibnamefont {Wang}}, \bibinfo
  {author} {\bibfnamefont {J.~S.}\ \bibnamefont {Wen}}, \bibinfo {author}
  {\bibfnamefont {Z.~J.}\ \bibnamefont {Xu}}, \bibinfo {author} {\bibfnamefont
  {G.}~\bibnamefont {Gu}}, \bibinfo {author} {\bibfnamefont {Y.}~\bibnamefont
  {Yoshida}}, \bibinfo {author} {\bibfnamefont {H.}~\bibnamefont {Eisaki}},
  \bibinfo {author} {\bibfnamefont {G.~B.}\ \bibnamefont {Arnold}},\ and\
  \bibinfo {author} {\bibfnamefont {D.~S.}\ \bibnamefont {Dessau}},\ }\bibfield
   {title} {\bibinfo {title} {{A unified form of low-energy nodal electronic
  interactions in hole-doped cuprate superconductors}},\ }\href
  {https://doi.org/10.1038/s41467-019-13497-4} {\bibfield  {journal} {\bibinfo
  {journal} {Nature Communications}\ }\textbf {\bibinfo {volume} {10}},\
  \bibinfo {pages} {5737} (\bibinfo {year} {2019})},\ \Eprint
  {https://arxiv.org/abs/1509.01611} {1509.01611} \BibitemShut {NoStop}%
\bibitem [{\citenamefont {Hartnoll}\ \emph {et~al.}(2016)\citenamefont
  {Hartnoll}, \citenamefont {Lucas},\ and\ \citenamefont
  {Sachdev}}]{hartnoll2016holographic}%
  \BibitemOpen
  \bibfield  {author} {\bibinfo {author} {\bibfnamefont {S.~A.}\ \bibnamefont
  {Hartnoll}}, \bibinfo {author} {\bibfnamefont {A.}~\bibnamefont {Lucas}},\
  and\ \bibinfo {author} {\bibfnamefont {S.}~\bibnamefont {Sachdev}},\
  }\href@noop {} {\emph {\bibinfo {title} {Holographic quantum matter}}}\
  (\bibinfo  {publisher} {The MIT Press},\ \bibinfo {year} {2016})\BibitemShut
  {NoStop}%
\bibitem [{\citenamefont {Zaanen}\ \emph {et~al.}(2015)\citenamefont {Zaanen},
  \citenamefont {Liu}, \citenamefont {Sun},\ and\ \citenamefont
  {Schalm}}]{Zaanen2015}%
  \BibitemOpen
  \bibfield  {author} {\bibinfo {author} {\bibfnamefont {J.}~\bibnamefont
  {Zaanen}}, \bibinfo {author} {\bibfnamefont {Y.}~\bibnamefont {Liu}},
  \bibinfo {author} {\bibfnamefont {Y.-W.}\ \bibnamefont {Sun}},\ and\ \bibinfo
  {author} {\bibfnamefont {K.}~\bibnamefont {Schalm}},\ }\href
  {https://doi.org/10.1017/cbo9781139942492} {\emph {\bibinfo {title}
  {Holographic Duality in Condensed Matter Physics}}}\ (\bibinfo  {publisher}
  {Cambridge University Press},\ \bibinfo {year} {2015})\BibitemShut {NoStop}%
\bibitem [{\citenamefont {Cubrovic}\ \emph {et~al.}(2009)\citenamefont
  {Cubrovic}, \citenamefont {Zaanen},\ and\ \citenamefont
  {Schalm}}]{Cubrovic2009}%
  \BibitemOpen
  \bibfield  {author} {\bibinfo {author} {\bibfnamefont {M.}~\bibnamefont
  {Cubrovic}}, \bibinfo {author} {\bibfnamefont {J.}~\bibnamefont {Zaanen}},\
  and\ \bibinfo {author} {\bibfnamefont {K.}~\bibnamefont {Schalm}},\
  }\bibfield  {title} {\bibinfo {title} {{String Theory, Quantum Phase
  Transitions, and the Emergent Fermi Liquid}},\ }\href@noop {} {\bibfield
  {journal} {\bibinfo  {journal} {Science}\ }\textbf {\bibinfo {volume}
  {325}},\ \bibinfo {pages} {439} (\bibinfo {year} {2009})}\BibitemShut
  {NoStop}%
\bibitem [{\citenamefont {Faulkner}\ \emph {et~al.}(2010)\citenamefont
  {Faulkner}, \citenamefont {Iqbal}, \citenamefont {Liu}, \citenamefont
  {McGreevy},\ and\ \citenamefont {Vegh}}]{Faulkner2010}%
  \BibitemOpen
  \bibfield  {author} {\bibinfo {author} {\bibfnamefont {T.}~\bibnamefont
  {Faulkner}}, \bibinfo {author} {\bibfnamefont {N.}~\bibnamefont {Iqbal}},
  \bibinfo {author} {\bibfnamefont {H.}~\bibnamefont {Liu}}, \bibinfo {author}
  {\bibfnamefont {J.}~\bibnamefont {McGreevy}},\ and\ \bibinfo {author}
  {\bibfnamefont {D.}~\bibnamefont {Vegh}},\ }\bibfield  {title} {\bibinfo
  {title} {{Strange Metal Transport Realized by Gauge/Gravity Duality}},\
  }\href@noop {} {\bibfield  {journal} {\bibinfo  {journal} {Science}\ }\textbf
  {\bibinfo {volume} {329}},\ \bibinfo {pages} {1043} (\bibinfo {year}
  {2010})}\BibitemShut {NoStop}%
\bibitem [{\citenamefont {Liu}\ \emph {et~al.}(2011)\citenamefont {Liu},
  \citenamefont {McGreevy},\ and\ \citenamefont {Vegh}}]{Liu2011}%
  \BibitemOpen
  \bibfield  {author} {\bibinfo {author} {\bibfnamefont {H.}~\bibnamefont
  {Liu}}, \bibinfo {author} {\bibfnamefont {J.}~\bibnamefont {McGreevy}},\ and\
  \bibinfo {author} {\bibfnamefont {D.}~\bibnamefont {Vegh}},\ }\bibfield
  {title} {\bibinfo {title} {Non-fermi liquids from holography},\ }\href
  {https://doi.org/10.1103/PhysRevD.83.065029} {\bibfield  {journal} {\bibinfo
  {journal} {Phys. Rev. D}\ }\textbf {\bibinfo {volume} {83}},\ \bibinfo
  {pages} {065029} (\bibinfo {year} {2011})}\BibitemShut {NoStop}%
\bibitem [{\citenamefont {Leong}\ \emph {et~al.}(2017)\citenamefont {Leong},
  \citenamefont {Setty}, \citenamefont {Limtragool},\ and\ \citenamefont
  {Phillips}}]{Leong2017}%
  \BibitemOpen
  \bibfield  {author} {\bibinfo {author} {\bibfnamefont {Z.}~\bibnamefont
  {Leong}}, \bibinfo {author} {\bibfnamefont {C.}~\bibnamefont {Setty}},
  \bibinfo {author} {\bibfnamefont {K.}~\bibnamefont {Limtragool}},\ and\
  \bibinfo {author} {\bibfnamefont {P.~W.}\ \bibnamefont {Phillips}},\
  }\bibfield  {title} {\bibinfo {title} {{Power-law liquid in cuprate
  superconductors from fermionic unparticles}},\ }\href@noop {} {\bibfield
  {journal} {\bibinfo  {journal} {Physical Review B}\ }\textbf {\bibinfo
  {volume} {96}},\ \bibinfo {pages} {1} (\bibinfo {year} {2017})}\BibitemShut
  {NoStop}%
\bibitem [{\citenamefont {Faulkner}\ and\ \citenamefont
  {Polchinski}(2011)}]{Faulkner:2010tq}%
  \BibitemOpen
  \bibfield  {author} {\bibinfo {author} {\bibfnamefont {T.}~\bibnamefont
  {Faulkner}}\ and\ \bibinfo {author} {\bibfnamefont {J.}~\bibnamefont
  {Polchinski}},\ }\bibfield  {title} {\bibinfo {title} {{Semi-Holographic
  Fermi Liquids}},\ }\href {https://doi.org/10.1007/JHEP06(2011)012} {\bibfield
   {journal} {\bibinfo  {journal} {Journal of High Energy Physics}\ }\textbf
  {\bibinfo {volume} {06}},\ \bibinfo {pages} {012} (\bibinfo {year} {2011})},\
  \Eprint {https://arxiv.org/abs/1001.5049} {1001.5049} \BibitemShut {NoStop}%
\bibitem [{\citenamefont {Bogdanov}\ \emph {et~al.}(2000)\citenamefont
  {Bogdanov}, \citenamefont {Lanzara}, \citenamefont {Kellar}, \citenamefont
  {Zhou}, \citenamefont {Lu}, \citenamefont {Zheng}, \citenamefont {Gu},
  \citenamefont {Shimoyama}, \citenamefont {Kishio}, \citenamefont {Ikeda},
  \citenamefont {Yoshizaki}, \citenamefont {Hussain},\ and\ \citenamefont
  {Shen}}]{Bogdanov2000}%
  \BibitemOpen
  \bibfield  {author} {\bibinfo {author} {\bibfnamefont {P.~V.}\ \bibnamefont
  {Bogdanov}}, \bibinfo {author} {\bibfnamefont {A.}~\bibnamefont {Lanzara}},
  \bibinfo {author} {\bibfnamefont {S.~A.}\ \bibnamefont {Kellar}}, \bibinfo
  {author} {\bibfnamefont {X.~J.}\ \bibnamefont {Zhou}}, \bibinfo {author}
  {\bibfnamefont {E.~D.}\ \bibnamefont {Lu}}, \bibinfo {author} {\bibfnamefont
  {W.~J.}\ \bibnamefont {Zheng}}, \bibinfo {author} {\bibfnamefont
  {G.}~\bibnamefont {Gu}}, \bibinfo {author} {\bibfnamefont {J.~I.}\
  \bibnamefont {Shimoyama}}, \bibinfo {author} {\bibfnamefont {K.}~\bibnamefont
  {Kishio}}, \bibinfo {author} {\bibfnamefont {H.}~\bibnamefont {Ikeda}},
  \bibinfo {author} {\bibfnamefont {R.}~\bibnamefont {Yoshizaki}}, \bibinfo
  {author} {\bibfnamefont {Z.}~\bibnamefont {Hussain}},\ and\ \bibinfo {author}
  {\bibfnamefont {Z.~X.}\ \bibnamefont {Shen}},\ }\bibfield  {title} {\bibinfo
  {title} {{Evidence for an energy scale for quasiparticle dispersion in
  Bi$_{2}$Sr$_{2}$CaCu$_{2}$O$_{8+ \delta}$}},\ }\href@noop {} {\bibfield
  {journal} {\bibinfo  {journal} {Physical Review Letters}\ }\textbf {\bibinfo
  {volume} {85}},\ \bibinfo {pages} {2581} (\bibinfo {year}
  {2000})}\BibitemShut {NoStop}%
\bibitem [{\citenamefont {Cuk}\ \emph {et~al.}(2005)\citenamefont {Cuk},
  \citenamefont {Lu}, \citenamefont {Zhou}, \citenamefont {Shen}, \citenamefont
  {Devereaux},\ and\ \citenamefont {Nagaosa}}]{Cuk2005}%
  \BibitemOpen
  \bibfield  {author} {\bibinfo {author} {\bibfnamefont {T.}~\bibnamefont
  {Cuk}}, \bibinfo {author} {\bibfnamefont {D.~H.}\ \bibnamefont {Lu}},
  \bibinfo {author} {\bibfnamefont {X.~J.}\ \bibnamefont {Zhou}}, \bibinfo
  {author} {\bibfnamefont {Z.~X.}\ \bibnamefont {Shen}}, \bibinfo {author}
  {\bibfnamefont {T.~P.}\ \bibnamefont {Devereaux}},\ and\ \bibinfo {author}
  {\bibfnamefont {N.}~\bibnamefont {Nagaosa}},\ }\bibfield  {title} {\bibinfo
  {title} {{A review of electron-phonon coupling seen in the high-Tc.
  Superconductors by angle-resolved photoemission studies (ARPES)}},\
  }\href@noop {} {\bibfield  {journal} {\bibinfo  {journal} {Physica Status
  Solidi (B) Basic Research}\ }\textbf {\bibinfo {volume} {242}},\ \bibinfo
  {pages} {11} (\bibinfo {year} {2005})}\BibitemShut {NoStop}%
\bibitem [{\citenamefont {Kaminski}\ and\ \citenamefont
  {Fretwell}(2004)}]{Gaebel2004}%
  \BibitemOpen
  \bibfield  {author} {\bibinfo {author} {\bibfnamefont {A.}~\bibnamefont
  {Kaminski}}\ and\ \bibinfo {author} {\bibfnamefont {H.~M.}\ \bibnamefont
  {Fretwell}},\ }\bibfield  {title} {\bibinfo {title} {{On the extraction of
  the self-energy from angle- resolved photoemission spectroscopy}},\
  }\href@noop {} {\bibfield  {journal} {\bibinfo  {journal} {New Journal of
  Physics}\ }\textbf {\bibinfo {volume} {6}},\ \bibinfo {pages} {1} (\bibinfo
  {year} {2004})}\BibitemShut {NoStop}%
\bibitem [{\citenamefont {Faulkner}\ \emph
  {et~al.}(2011{\natexlab{a}})\citenamefont {Faulkner}, \citenamefont {Liu},
  \citenamefont {McGreevy},\ and\ \citenamefont {Vegh}}]{Faulkner2011a}%
  \BibitemOpen
  \bibfield  {author} {\bibinfo {author} {\bibfnamefont {T.}~\bibnamefont
  {Faulkner}}, \bibinfo {author} {\bibfnamefont {H.}~\bibnamefont {Liu}},
  \bibinfo {author} {\bibfnamefont {J.}~\bibnamefont {McGreevy}},\ and\
  \bibinfo {author} {\bibfnamefont {D.}~\bibnamefont {Vegh}},\ }\bibfield
  {title} {\bibinfo {title} {{Emergent quantum criticality, Fermi surfaces, and
  AdS$_2$}},\ }\href@noop {} {\bibfield  {journal} {\bibinfo  {journal}
  {Physical Review D}\ }\textbf {\bibinfo {volume} {83}},\ \bibinfo {pages}
  {125002} (\bibinfo {year} {2011}{\natexlab{a}})}\BibitemShut {NoStop}%
\bibitem [{\citenamefont {Gubser}\ and\ \citenamefont
  {Rocha}(2010)}]{Gubser2010}%
  \BibitemOpen
  \bibfield  {author} {\bibinfo {author} {\bibfnamefont {S.~S.}\ \bibnamefont
  {Gubser}}\ and\ \bibinfo {author} {\bibfnamefont {F.~D.}\ \bibnamefont
  {Rocha}},\ }\bibfield  {title} {\bibinfo {title} {{Peculiar properties of a
  charged dilatonic black hole in AdS$_5$}},\ }\href@noop {} {\bibfield
  {journal} {\bibinfo  {journal} {Physical Review D}\ }\textbf {\bibinfo
  {volume} {81}},\ \bibinfo {pages} {1} (\bibinfo {year} {2010})}\BibitemShut
  {NoStop}%
\bibitem [{\citenamefont {Gubser}\ and\ \citenamefont
  {Ren}(2012)}]{Gubser2012}%
  \BibitemOpen
  \bibfield  {author} {\bibinfo {author} {\bibfnamefont {S.~S.}\ \bibnamefont
  {Gubser}}\ and\ \bibinfo {author} {\bibfnamefont {J.}~\bibnamefont {Ren}},\
  }\bibfield  {title} {\bibinfo {title} {Analytic fermionic green's functions
  from holography},\ }\href {https://doi.org/10.1103/PhysRevD.86.046004}
  {\bibfield  {journal} {\bibinfo  {journal} {Phys. Rev. D}\ }\textbf {\bibinfo
  {volume} {86}},\ \bibinfo {pages} {046004} (\bibinfo {year}
  {2012})}\BibitemShut {NoStop}%
\bibitem [{\citenamefont {Randeria}\ \emph {et~al.}(2004)\citenamefont
  {Randeria}, \citenamefont {Paramekanti},\ and\ \citenamefont
  {Trivedi}}]{Randeria2004}%
  \BibitemOpen
  \bibfield  {author} {\bibinfo {author} {\bibfnamefont {M.}~\bibnamefont
  {Randeria}}, \bibinfo {author} {\bibfnamefont {A.}~\bibnamefont
  {Paramekanti}},\ and\ \bibinfo {author} {\bibfnamefont {N.}~\bibnamefont
  {Trivedi}},\ }\bibfield  {title} {\bibinfo {title} {{Nodal quasiparticle
  dispersion in strongly correlated d-wave superconductors}},\ }\href
  {https://doi.org/10.1103/PhysRevB.69.144509} {\bibfield  {journal} {\bibinfo
  {journal} {Physical Review B - Condensed Matter and Materials Physics}\
  }\textbf {\bibinfo {volume} {69}},\ \bibinfo {pages} {1} (\bibinfo {year}
  {2004})}\BibitemShut {NoStop}%
\bibitem [{\citenamefont {Veenstra}\ \emph {et~al.}(2011)\citenamefont
  {Veenstra}, \citenamefont {Goodvin}, \citenamefont {Berciu},\ and\
  \citenamefont {Damascelli}}]{Veenstra2011}%
  \BibitemOpen
  \bibfield  {author} {\bibinfo {author} {\bibfnamefont {C.~N.}\ \bibnamefont
  {Veenstra}}, \bibinfo {author} {\bibfnamefont {G.~L.}\ \bibnamefont
  {Goodvin}}, \bibinfo {author} {\bibfnamefont {M.}~\bibnamefont {Berciu}},\
  and\ \bibinfo {author} {\bibfnamefont {A.}~\bibnamefont {Damascelli}},\
  }\bibfield  {title} {\bibinfo {title} {{Spectral function tour of
  electron-phonon coupling outside the Migdal limit}},\ }\href
  {https://doi.org/10.1103/PhysRevB.84.085126} {\bibfield  {journal} {\bibinfo
  {journal} {Physical Review B - Condensed Matter and Materials Physics}\
  }\textbf {\bibinfo {volume} {84}},\ \bibinfo {pages} {21} (\bibinfo {year}
  {2011})}\BibitemShut {NoStop}%
\bibitem [{\citenamefont {Jin}\ \emph {et~al.}(2019)\citenamefont {Jin},
  \citenamefont {Tsyplyatyev}, \citenamefont {Moreno}, \citenamefont {Anthore},
  \citenamefont {Tan}, \citenamefont {Griffiths}, \citenamefont {Farrer},
  \citenamefont {Ritchie}, \citenamefont {Glazman}, \citenamefont {Schofield},\
  and\ \citenamefont {Ford}}]{Jin2019}%
  \BibitemOpen
  \bibfield  {author} {\bibinfo {author} {\bibfnamefont {Y.}~\bibnamefont
  {Jin}}, \bibinfo {author} {\bibfnamefont {O.}~\bibnamefont {Tsyplyatyev}},
  \bibinfo {author} {\bibfnamefont {M.}~\bibnamefont {Moreno}}, \bibinfo
  {author} {\bibfnamefont {A.}~\bibnamefont {Anthore}}, \bibinfo {author}
  {\bibfnamefont {W.~K.}\ \bibnamefont {Tan}}, \bibinfo {author} {\bibfnamefont
  {J.~P.}\ \bibnamefont {Griffiths}}, \bibinfo {author} {\bibfnamefont
  {I.}~\bibnamefont {Farrer}}, \bibinfo {author} {\bibfnamefont {D.~A.}\
  \bibnamefont {Ritchie}}, \bibinfo {author} {\bibfnamefont {L.~I.}\
  \bibnamefont {Glazman}}, \bibinfo {author} {\bibfnamefont {A.~J.}\
  \bibnamefont {Schofield}},\ and\ \bibinfo {author} {\bibfnamefont {C.~J.}\
  \bibnamefont {Ford}},\ }\bibfield  {title} {\bibinfo {title}
  {{Momentum-dependent power law measured in an interacting quantum wire beyond
  the Luttinger limit}},\ }\href@noop {} {\bibfield  {journal} {\bibinfo
  {journal} {Nature Communications}\ }\textbf {\bibinfo {volume} {10}}
  (\bibinfo {year} {2019})}\BibitemShut {NoStop}%
\bibitem [{\citenamefont {Imambekov}\ and\ \citenamefont
  {Glazman}(2009)}]{Imambekov2009}%
  \BibitemOpen
  \bibfield  {author} {\bibinfo {author} {\bibfnamefont {A.}~\bibnamefont
  {Imambekov}}\ and\ \bibinfo {author} {\bibfnamefont {L.}~\bibnamefont
  {Glazman}},\ }\bibfield  {title} {\bibinfo {title} {Universal theory of
  nonlinear luttinger liquids},\ }\href
  {https://doi.org/doi.org/10.1126/science.1165403} {\bibfield  {journal}
  {\bibinfo  {journal} {Science}\ }\textbf {\bibinfo {volume} {323}},\ \bibinfo
  {pages} {228} (\bibinfo {year} {2009})}\BibitemShut {NoStop}%
\bibitem [{\citenamefont {Jacobs}\ \emph {et~al.}(2015)\citenamefont {Jacobs},
  \citenamefont {Grubinskas},\ and\ \citenamefont {Stoof}}]{Jacobs2015}%
  \BibitemOpen
  \bibfield  {author} {\bibinfo {author} {\bibfnamefont {V.~P.~J.}\
  \bibnamefont {Jacobs}}, \bibinfo {author} {\bibfnamefont {S.}~\bibnamefont
  {Grubinskas}},\ and\ \bibinfo {author} {\bibfnamefont {H.~T.~C.}\
  \bibnamefont {Stoof}},\ }\bibfield  {title} {\bibinfo {title} {{Towards a
  field-theory interpretation of bottom-up holography}},\ }\href
  {https://doi.org/10.1007/JHEP04(2015)033} {\bibfield  {journal} {\bibinfo
  {journal} {Journal of High Energy Physics}\ }\textbf {\bibinfo {volume}
  {2015}},\ \bibinfo {pages} {33} (\bibinfo {year} {2015})}\BibitemShut
  {NoStop}%
\bibitem [{\citenamefont {Reber}\ \emph {et~al.}(2014)\citenamefont {Reber},
  \citenamefont {Plumb}, \citenamefont {Waugh},\ and\ \citenamefont
  {Dessau}}]{Reber2014}%
  \BibitemOpen
  \bibfield  {author} {\bibinfo {author} {\bibfnamefont {T.~J.}\ \bibnamefont
  {Reber}}, \bibinfo {author} {\bibfnamefont {N.~C.}\ \bibnamefont {Plumb}},
  \bibinfo {author} {\bibfnamefont {J.~A.}\ \bibnamefont {Waugh}},\ and\
  \bibinfo {author} {\bibfnamefont {D.~S.}\ \bibnamefont {Dessau}},\ }\bibfield
   {title} {\bibinfo {title} {{Effects, determination, and correction of count
  rate nonlinearity in multi-channel analog electron detectors}},\ }\href@noop
  {} {\bibfield  {journal} {\bibinfo  {journal} {Review of Scientific
  Instruments}\ }\textbf {\bibinfo {volume} {85}},\ \bibinfo {pages} {043907}
  (\bibinfo {year} {2014})}\BibitemShut {NoStop}%
\bibitem [{\citenamefont {Presland}\ \emph {et~al.}(1991)\citenamefont
  {Presland}, \citenamefont {Tallon}, \citenamefont {Buckley}, \citenamefont
  {Liu},\ and\ \citenamefont {Flower}}]{Presland1991}%
  \BibitemOpen
  \bibfield  {author} {\bibinfo {author} {\bibfnamefont {M.~R.}\ \bibnamefont
  {Presland}}, \bibinfo {author} {\bibfnamefont {J.~L.}\ \bibnamefont
  {Tallon}}, \bibinfo {author} {\bibfnamefont {R.~G.}\ \bibnamefont {Buckley}},
  \bibinfo {author} {\bibfnamefont {R.~S.}\ \bibnamefont {Liu}},\ and\ \bibinfo
  {author} {\bibfnamefont {N.~E.}\ \bibnamefont {Flower}},\ }\bibfield  {title}
  {\bibinfo {title} {{General trends in oxygen stoichiometry effects on $T_c$
  in Bi and Tl superconductors}},\ }\href
  {https://doi.org/10.1016/0921-4534(91)90700-9} {\bibfield  {journal}
  {\bibinfo  {journal} {Physica C: Superconductivity and its applications}\
  }\textbf {\bibinfo {volume} {176}},\ \bibinfo {pages} {95} (\bibinfo {year}
  {1991})}\BibitemShut {NoStop}%
\bibitem [{\citenamefont {Ando}\ \emph {et~al.}(2000)\citenamefont {Ando},
  \citenamefont {Hanaki}, \citenamefont {Ono}, \citenamefont {Murayama},
  \citenamefont {Segawa}, \citenamefont {Miyamoto},\ and\ \citenamefont
  {Komiya}}]{ando2000}%
  \BibitemOpen
  \bibfield  {author} {\bibinfo {author} {\bibfnamefont {Y.}~\bibnamefont
  {Ando}}, \bibinfo {author} {\bibfnamefont {Y.}~\bibnamefont {Hanaki}},
  \bibinfo {author} {\bibfnamefont {S.}~\bibnamefont {Ono}}, \bibinfo {author}
  {\bibfnamefont {T.}~\bibnamefont {Murayama}}, \bibinfo {author}
  {\bibfnamefont {K.}~\bibnamefont {Segawa}}, \bibinfo {author} {\bibfnamefont
  {N.}~\bibnamefont {Miyamoto}},\ and\ \bibinfo {author} {\bibfnamefont
  {S.}~\bibnamefont {Komiya}},\ }\bibfield  {title} {\bibinfo {title} {{Carrier
  concentrations in Bi$_2$Sr$_{2-x}$La$_{x}$CuO$_{6+\delta}$ single crystals
  and their relation to the Hall coefficient and thermopower}},\ }\href
  {https://doi.org/10.1103/PhysRevB.61.R14956} {\bibfield  {journal} {\bibinfo
  {journal} {Physical Review B}\ }\textbf {\bibinfo {volume} {61}},\ \bibinfo
  {pages} {R14956} (\bibinfo {year} {2000})},\ \Eprint
  {https://arxiv.org/abs/0004133} {0004133} \BibitemShut {NoStop}%
\bibitem [{\citenamefont {G{\"{u}}rsoy}\ \emph {et~al.}(2012)\citenamefont
  {G{\"{u}}rsoy}, \citenamefont {Plauschinn}, \citenamefont {Stoof},\ and\
  \citenamefont {Vandoren}}]{Gursoy2012}%
  \BibitemOpen
  \bibfield  {author} {\bibinfo {author} {\bibfnamefont {U.}~\bibnamefont
  {G{\"{u}}rsoy}}, \bibinfo {author} {\bibfnamefont {E.}~\bibnamefont
  {Plauschinn}}, \bibinfo {author} {\bibfnamefont {H.}~\bibnamefont {Stoof}},\
  and\ \bibinfo {author} {\bibfnamefont {S.}~\bibnamefont {Vandoren}},\
  }\bibfield  {title} {\bibinfo {title} {{Holography and ARPES sum-rules}},\
  }\href {https://doi.org/10.1007/JHEP05(2012)018} {\bibfield  {journal}
  {\bibinfo  {journal} {Journal of High Energy Physics}\ }\textbf {\bibinfo
  {volume} {2012}},\ \bibinfo {pages} {18} (\bibinfo {year} {2012})},\ \Eprint
  {https://arxiv.org/abs/1112.5074} {1112.5074} \BibitemShut {NoStop}%
\bibitem [{\citenamefont {Iqbal}\ \emph {et~al.}(2011)\citenamefont {Iqbal},
  \citenamefont {Liu},\ and\ \citenamefont {Mezei}}]{Iqbal2011}%
  \BibitemOpen
  \bibfield  {author} {\bibinfo {author} {\bibfnamefont {N.}~\bibnamefont
  {Iqbal}}, \bibinfo {author} {\bibfnamefont {H.}~\bibnamefont {Liu}},\ and\
  \bibinfo {author} {\bibfnamefont {M.}~\bibnamefont {Mezei}},\ }\bibinfo
  {title} {Lectures on holographic non-fermi liquids and quantum phase
  transitions},\ in\ \href {https://doi.org/10.1142/9789814350525_0013} {\emph
  {\bibinfo {booktitle} {String Theory and Its Applications}}}\ (\bibinfo
  {year} {2011})\ pp.\ \bibinfo {pages} {707--815}\BibitemShut {NoStop}%
\bibitem [{\citenamefont {Faulkner}\ \emph
  {et~al.}(2011{\natexlab{b}})\citenamefont {Faulkner}, \citenamefont {Liu},\
  and\ \citenamefont {Rangamani}}]{Faulkner2011}%
  \BibitemOpen
  \bibfield  {author} {\bibinfo {author} {\bibfnamefont {T.}~\bibnamefont
  {Faulkner}}, \bibinfo {author} {\bibfnamefont {H.}~\bibnamefont {Liu}},\ and\
  \bibinfo {author} {\bibfnamefont {M.}~\bibnamefont {Rangamani}},\ }\bibfield
  {title} {\bibinfo {title} {{Integrating out geometry: holographic Wilsonian
  RG and the membrane paradigm}},\ }\href
  {https://doi.org/10.1007/JHEP08(2011)051} {\bibfield  {journal} {\bibinfo
  {journal} {Journal of High Energy Physics}\ }\textbf {\bibinfo {volume}
  {2011}},\ \bibinfo {pages} {51} (\bibinfo {year}
  {2011}{\natexlab{b}})}\BibitemShut {NoStop}%
\end{thebibliography}%

\section*{Acknowledgements}
We acknowledge the former Foundation for Fundamental Research on Matter (FOM) under grant no. 16METL01, ‘Strange Metals’. FOM was financially supported by the Netherlands Organization for Scientific Research (NWO), which in its turn is funded by the Dutch Ministry of Education, Culture and Science (OCW). This work is also part of the D-ITP consortium, a program of the NWO. 
We acknowledge Diamond Light Source for time on Beamline I05 under Proposals SI19403 and SI22464, and Jan de Boer, Stefan Vandoren, Jay Armas and Umut G{\"{u}}rsoy for a critical reading of the manuscript and we are grateful to Jan Zaanen for many useful discussions.

\end{document}